\documentclass[usenatbib]{mn2e}
\usepackage{epsf,wrapfig}
\usepackage{graphicx}
\usepackage{amssymb}
\usepackage{deluxetable}

\begin{document}

\title[A large-scale survey of X-ray filaments]{A Large Scale Survey
  of X-Ray Filaments in the Galactic Center}
\author[S.P. Johnson, H. Dong and Q. D. Wang]{S. P. Johnson$^1$, H. Dong,
  and Q. D. Wang\\
$^1$Department of Astronomy, University of Massachusetts Amherst,
  Amherst, MA, USA\\
\tt e-mail: spjohnso@astro.umass.edu}

\maketitle


\begin{abstract}

We present a catalog of 17 filamentary X-ray features located within a
68'$\times$34' view centered on the Galactic Center region from images taken by
\textit{Chandra}.  These features are described by their morphological
and spectral properties.  Many of the X-ray features have non-thermal
spectra that are well fit by an absorbed power-law.  Of the 17
features, we find 6 that have not been previously detected, 4 of which
are outside the immediate 20'$\times$20' area centered on the GC.
7 of the 17 identified filaments have morphological and
spectral properties expected for pulsar wind nebulae with X-ray
luminosities of 5$\times$10$^{32}$ to 10$^{34}$ ergs
s$^{-1}$ in the 2.0-10.0 keV band and photon indexes in the range of
$\Gamma$=1.1 to 1.9. In one feature, we suggest
the strong neutral Fe K$\alpha$ emission line to be a possible
indicator for past activity of Sgr A*. For G359.942-0.03, a particular
filament of interest, we propose the model of a ram-pressure confined
stellar wind bubble from a massive star to account for the morphology,
spectral shape and 6.7 keV He-like Fe emission detected.  We also
present a piecewise spectral analysis on two features of interest,
G0.13-0.11 and G359.89-0.08, to further examine their physical
interpretations.  This analysis favors the PWN scenario for these
features. 
\end{abstract}


\section{Introduction}

The Galactic Center (GC) region is a unique environment that is home
to numerous energetic processes.  Radio, infrared and X-ray
observations show structures that can only be distinguished in our own
galaxy due to its proximity.  The GC thus provides a
laboratory in which to probe the Galactic nuclear environment and the
interactions between star formation regions, the interstellar medium (ISM)
and the supermassive black hole Sgr A*.  These observations can aid in
understanding the nuclear environments of nearby galaxies.

Radio observations have detected prominent non-thermal radio filaments
that have been intensively studied \citep[e.g.][]{Yusef84}.
Many of the non-thermal radio filaments may come from milliGauss magnetic
fields illuminated by energetic particles.  These fields are seen to
be primarily perpendicular to the galactic plane and indicate the general
structure of the magnetic field around the GC.  The exact origins and
implications of the radio non-thermal filaments in relation to the GC
are still in debate.  

\textit{Chandra} observations have detected numerous large scale X-ray
structures from thermal to non-thermal filaments
\citep[e.g.][]{bamba02,lu08,muno08} primarily in the local
20'$\times$20' area centered on the GC.  The commonly accepted scenarios
for non-thermal thread-like X-ray features in the GC include pulsar
wind nebulae (PWNe) \citep[e.g.][]{wang93,wang06}, with the
filament-like structure produced through ram-pressure confinement or
strong magnetic fields, and magnetohydrodynamical shock fronts
from supernova remnants (SNRs) \citep[e.g][]{Yusef05}.  For the SNR case, the
elongation of features represent shock fronts in the ISM.  The average
non-thermal spectrum of a typical PWN is generally well fit by a power
law model with a photon index in the range of $\Gamma$=1.1-2.4
\citep{gotthelf02} and X-ray luminosities in the range
of $10^{32}$ to $10^{37}$ ergs s$^{-1}$ for the 0.2-10 keV energy band
\citep[e.g.][]{gaensler06,kaspi06}.  The PWN scenario is motivated
primarily by the similarities between the features seen in the GC and
X-ray features associated with known pulsars \citep[see
  also][]{karga08}.  Few cases of K$\alpha$
transitions from neutral iron (Fe) are proposed to stem from the reflection of
hard X-rays from external sources, such as the supermassive black hole
Sgr A* \citep[e.g.][]{koyama89}.  By analyzing these X-ray features,
one can in principle trace gas dynamics and magnetic fields in the GC region
\citep[e.g.][]{wang02} and examine the history of GC activity
\citep{koyama89}.

Previous broad scale studies of X-ray features focus on the immediate
20'$\times$20' area surrounding Sgr A* \citep[e.g.][]{muno08,lu08}.
Here, we present a wide scale study on 17 X-ray thread-like 
features.  The features are detected over a roughly 68'$\times$34' area of
the sky, relating to approximately 158 pc by 79 pc, around Sgr A* based on
all available observations.  While some of these features have been examined
in previous literature, the improved counting statistics of this study
allows for piecewise analysis that will be able to
detect trends in the spectra and therefore better constrain the
physical model \citep[see also][]{lu08}.  In addition to the piecewise
analysis of some features, we also present an additional physical model in
order to explain a feature with a Helium (He)-like Fe K$\alpha$ emission
line, comet-like morphology and steep spectrum.  For
consistency within this study, the spectral and morphological
characteristics were performed for all features listed. All physical
distance measurements assume an 8 kpc distance to the GC and all
errors, unless otherwise stated, are given to the 90\% confidence level. 

\section{Observations and Detection}

The \textit{Chandra} X-ray Observatory observed the GC with the Advanced
CCD Imaging Spectrometer imaging array (ACIS-I) for a total integrated 
exposure time of $\sim$2 Msec between 1999 and 2007 over 81 observations
\citep{muno09}. The array is composed of four individual CCDs that
operate together to provide a 17'$\times$17' view with sub-arcsecond
resolution at the center of the array.  The images were first
reprocessed using the standard \textit{Chandra} Interactive Analysis
of Observations \textsc{ciao} routines (version 3.4.0) and then combined by
celestial coordinates to produce a merged events file.

Initial identification of X-ray filaments comes from a careful visual
examination of the merged event image in the 2.0-8.0 keV band.
We examine the merged events file under a logarithmic scale with
different upper/lower limits and look for extended, linear structures.
From this initial identification, we removed features whose
morphologies and apparent extended structures are not well confined
with respect to the feature in question, leaving only those with
tightly correlated extended X-ray emission.  Within regions of strong
diffuse X-ray emission, e.g. the Sgr A* complex, we limited our
selection to well constrained extended X-ray emission regions with
size scales larger than 3"$\times$1".  These morphologically selected
features were culled once more after taking their individual spectra
(see below for details on spectral extraction) and rejecting them if
their spectra appear consistent with their local background; e.g. in
spectral shape.  Based on our selection criteria, these features are
thus X-ray bright, in respect to their local backgrounds, and have
well defined linear morphologies; however, due to the complexity of
X-ray emission around the GC and the uneven sensitivity of this
survey, this sample is far from complete and represents only a
fraction of the extended X-ray features indigenous to the GC.
Fig. 1 shows the composite intensity  map in the 2.0-8.0 keV
band with identified features labeled and indicated by their surrounding
ellipses.  Fig. 2 shows the inner 20'$\times$20' around Sgr A* with
features again labeled and constrained by their surrounding ellipses.
In extracting their source and background spectra, point sources and
other such possible contaminators were excluded from the count and
exposure maps.  We use local background subtraction for each filament due
to the widespread area and time integration over which the filaments
are found.  For those that are surrounded by strong diffuse
emission, e.g. within the Sgr A* complex, backgrounds were
selected to account for the immediate intensities. 

\begin{figure}
\centering
\includegraphics[width=0.6\textwidth]{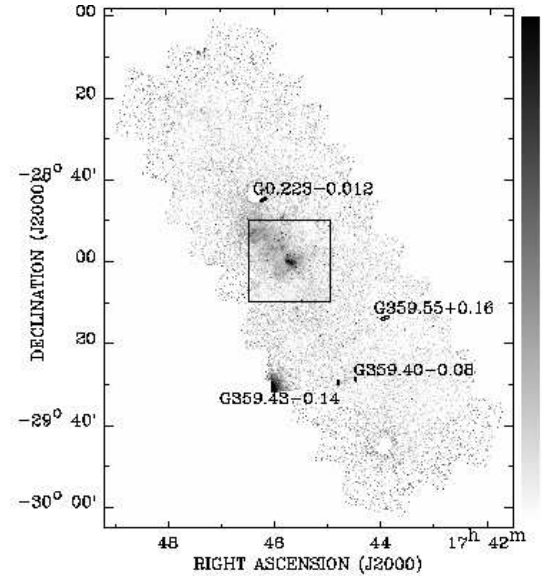}
\caption{Full view of the 81 compiled observations taken by
\textit{Chandra} around the GC.  The center square identifies the
20'$\times$20' region of Fig. 2.  Features outside this region are
selected by their surrounding ellipses and labeled based on their galactic
coordinates.  The image is the exposure-corrected intensity map in the
2.0-8.0 keV band produced by dividing the ACIS-I count map with the
exposure map.  This and subsequent images are given in logarithmic
scaling with North defined as up and East as left.}
\label{fig:fig1}
\end{figure}

\begin{figure}
\includegraphics[width=0.5\textwidth]{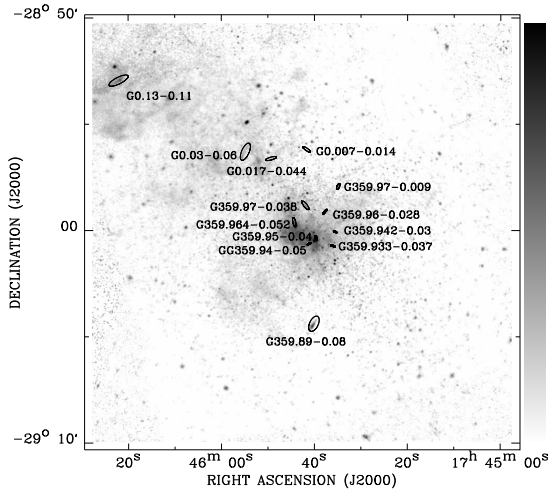}
\caption{Magnified view of the intensity map from Fig.1
  on the inner 20'$\times$20' area centered on Sgr A*.  Identified
  features are again selected and labeled by their surrounding ellipses.} 
\label{fig:fig2}
\end{figure}

Typical spectral extraction methods pose a problem for the  
features that were covered by multiple observations.
While some of the 81 observations have the same celestial coordinates
in their pointing, they could have different roll angles.  As such, a
feature could fall into different positions on the ACIS-I detector
over different observations including the gap between the CCDs that
compose the array.  Additionally, a feature may be completely
contained within one observation while partially covered in
another. Therefore, the traditional spectral extraction could not be
used due to time and spatial variability of the Auxiliary Response
File (\textsc{arf}) and Redistribution Matrix File (\textsc{rmf}) of
the instrument. Another way is to extract the spectra of the filaments
from individual observations and use "\textsc{ftools}" to merge them
into a final spectrum. However, this is just suitable for a single
object and can become difficult when dealing with many filaments over many
observations. Therefore, we developed a new method to directly extract
the spectrum from the merged events file. We first reproject all the
events in the merged events file of a certain filament into the
instrument coordinate, since the calculation of the \textsc{arf} and
\textsc{rmf} is based on events' positions on the CCD, not the celestial
coordinate. The \textsc{arf} is not only a function of the events' positions,
but also depends on when the observation was taken.  The quantum
efficiency of the CCDs has been seen to degrade over time in part due
to molecular contamination build up on the the CCD itself and/or the
filter.  However, we notice that above 2 keV, the \textsc{arf} of ACIS-I
decreases less than 5\% during the 7 years observational period
(absorption towards the GC is large for energies below 2 keV, making
the 2 keV lower limit optimal in terms of spectral analysis).  Using
this lower limit, the \textsc{arf} can then be considered to be
time-independent. The \textsc{rmf} file is very stable and is just a
function of position, not time (we have checked the \textsc{rmf} file
used by the \textsc{ciao} routines for each observation one by
one. All of them use the same calibration file). Therefore, a new
response matrix file, \textsc{rsp} (the multiple of the \textsc{arf}
and \textsc{rmf} file), for each small grid on the CCD (the place of
each grid in the CCD is predefined by the calibration file) with
events was created from the calibration file and the total number of
events falling into each grid was used as the weight for merging all
the new \textsc{rsp} files in different grids into the final merged
\textsc{rsp}.  We have tested this method in the center 7' region
around SGR A* and found that the fitting result from this method is
nearly identical to that from the same region in the observation with
the longest exposure time, proving that our method is robust and
reliable.

\section{Analysis of X-ray Features}

For each of the identified filaments, we classify its basic appearance
as well as spectral proprieties.  We broadly classify the features
into two morphological categories: 'filamentary' and 'cometary'.  A
feature is considered to be cometary if its morphology appears to
taper off in one direction, similar to a comet's tail, or if it has
apparent asymmetry in its brightness across its major axis.  Features
that do not meet these criteria are defined as filamentary, this
includes features that appear to be brightest in the center.  This
morphological classification along with the sizes of the features are
approximations based on the observed flux and apparent surface
brightness relative to the local background.  The name associated
with each feature comes from the approximate center or brightest point
of the feature. 

Fig. 3 through 7 show the individual features with ellipses
indicating regions used in extracting the source and background
spectra.  Linear X-ray features can be seen in some images in addition
to the features identified.  Many of these were either not selected
due to the possibility of being small scale structures of a larger
X-ray emitting region or rejected based on our selection criteria.
The plot next to each \textit{Chandra} count map gives the extracted
spectrum for each feature.  In each case, the spectrum is well fit
using a power law model with foreground absorption in \textsc{xspec}
(\textsc{pha(po)} for short).  A Gaussian model is included for those
features which exhibit line emission of 6.4 keV neutral Fe or 6.7 keV
He-like Fe (\textsc{pha(po+ga)}). The spectral parameters for the
\textsc{pha(po)} or \textsc{pha(po+ga)} single fit models as well as
the morphological properties for the features are listed in Tables 1
and 2.  It is well known that extinction towards the GC can vary
greatly over scales on order of arc-seconds
\citep[e.g.][]{scoville03}.  As such, each feature is modeled with an
independent value of column density ($N_{\rm{H}}$) rather than a single fixed
$N_{\rm{H}}$ for all features based on the assumed 8 kpc distance to the GC.
The following summary of features focuses on those with unique traits
or observations. 

\begin{figure*}
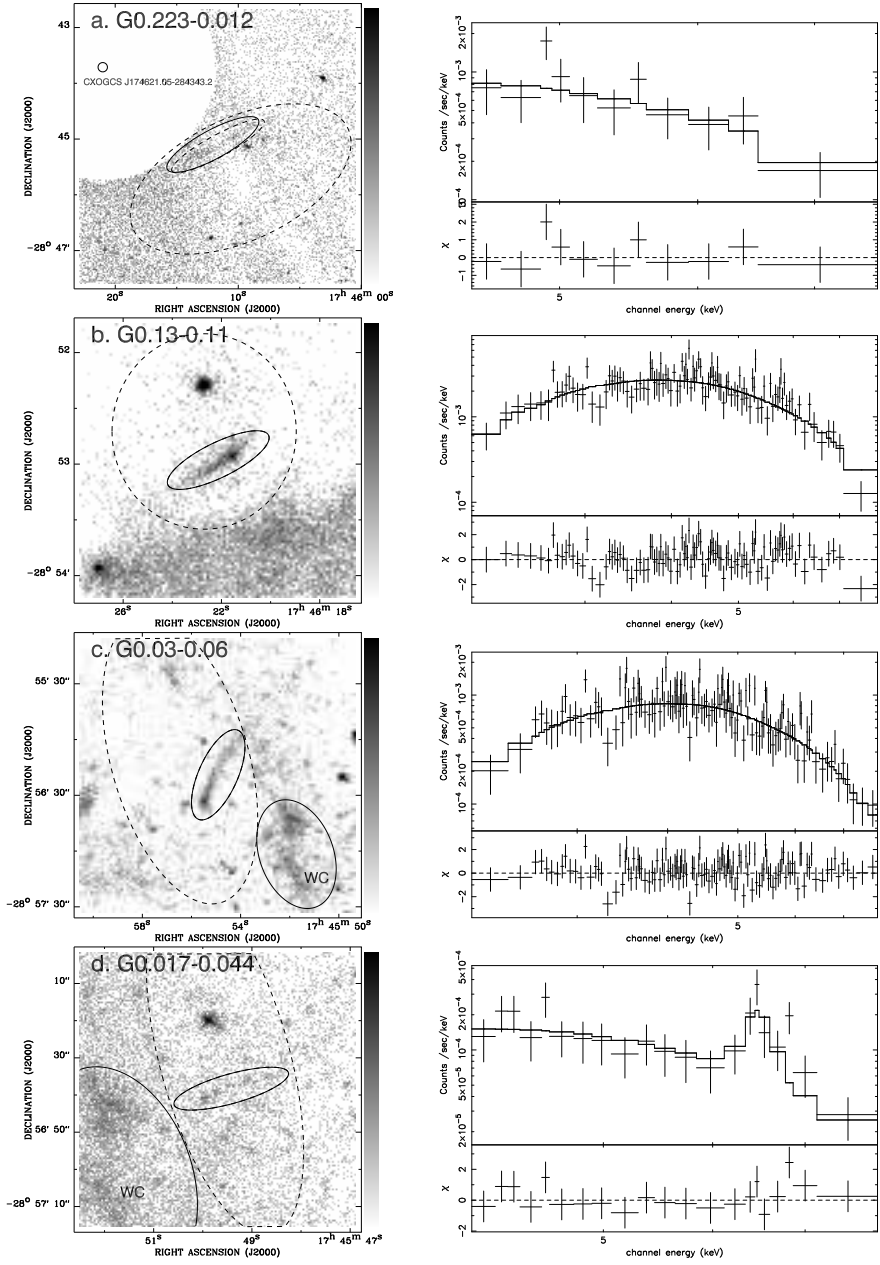

\include{fig3}
\caption[]{Left-hand panel: the X-ray count maps of individual
  features in the 2.0-8.0 keV band.  The upper and lower limits for
  each count map are 12 and 0.5 counts/arcsec$^2$ for
  G0.223-0.012 (a), 23 and 2.3 counts/arcsec$^2$ for
  G0.13-0.11 (b), 69 and 6.9 counts/arcsec$^2$ for
  G0.03-0.06 (c), and 207 and 4.1 counts/arcsec$^2$ for
  G0.017-0.044 (d), respectively.  Source regions are identified by the
  solid ellipses with background regions enclosed by dashed
  ellipses.  In 3a, the empty region results from the removal of
  events around the bright X-ray point source CXOGCS
  J174621.05-284343.2 whose position is given by the small circular
  region.  Fig. 3c and 3d include the 'West Clump' region from
  \cite{lu08}, labeled as 'WC'. Right-hand panel: the X-ray
  spectra for the matching features given in the left column in the
  2.0-8.0 keV band.  All spectra are fit with an absorbed power-law model
  (\textsc{pha(po)}). The model for G0.017-0.044 is fitted with a Gaussian
  centered at 6.4 keV in addition to the \textsc{pha(po)}}
\label{fig:fig3}
\end{figure*}

\begin{figure*}
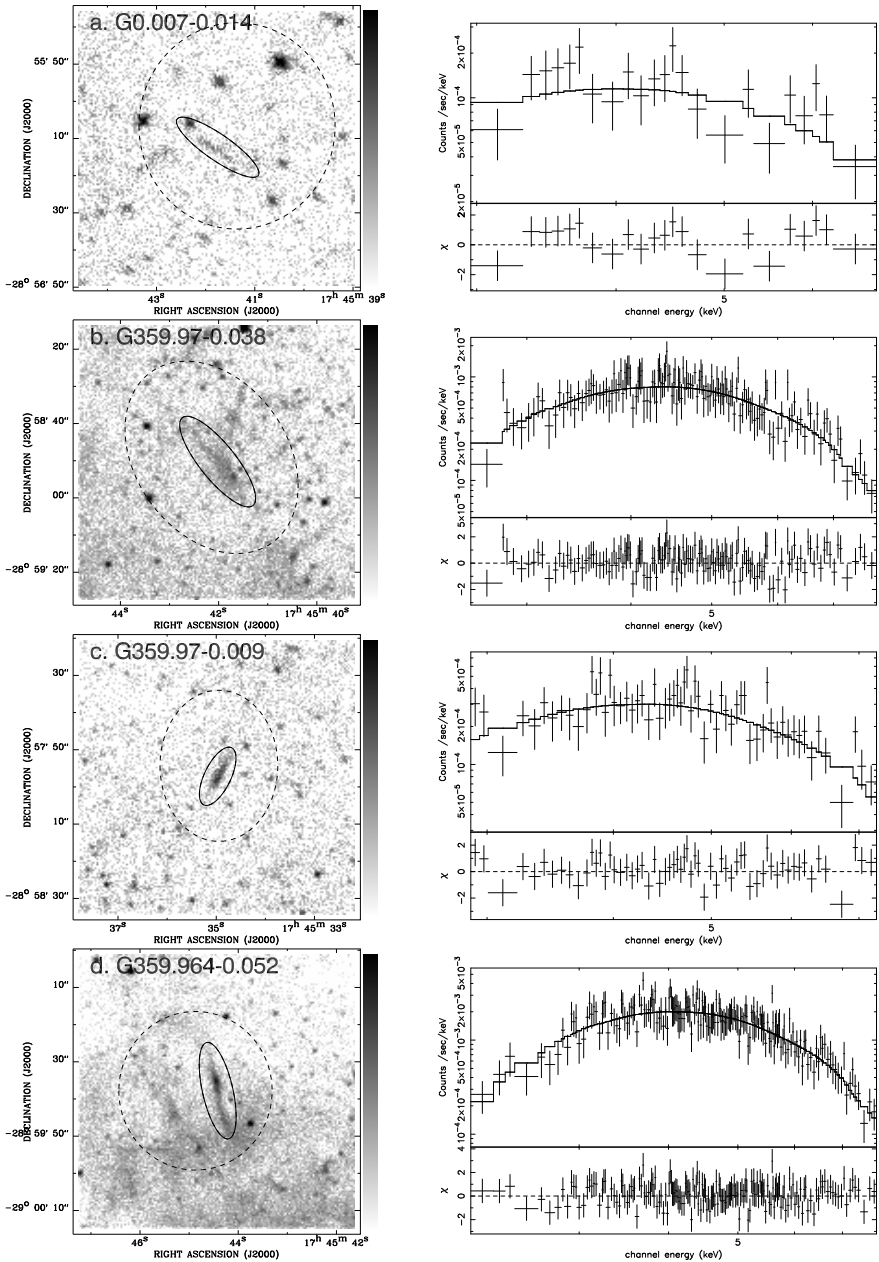

\include{fig4}
\caption{Continued X-ray count maps and spectra similar to Fig. 3.
  Upper and lower limits for each count map are 103 and 4.1
  counts/arcsec$^2$ for G0.007-0.014 (a), 410 and 4.1 counts/arcsec$^2$
  for G359.97-0.038 (b), 207 and 4.1 counts/arcsec$^2$ for G359.97-0.009
  (c), and 410 and 12 counts/arcsec$^2$ for G359.964-0.052 (d), respectively.} 
\label{fig:fig4}
\end{figure*}

\begin{figure*}
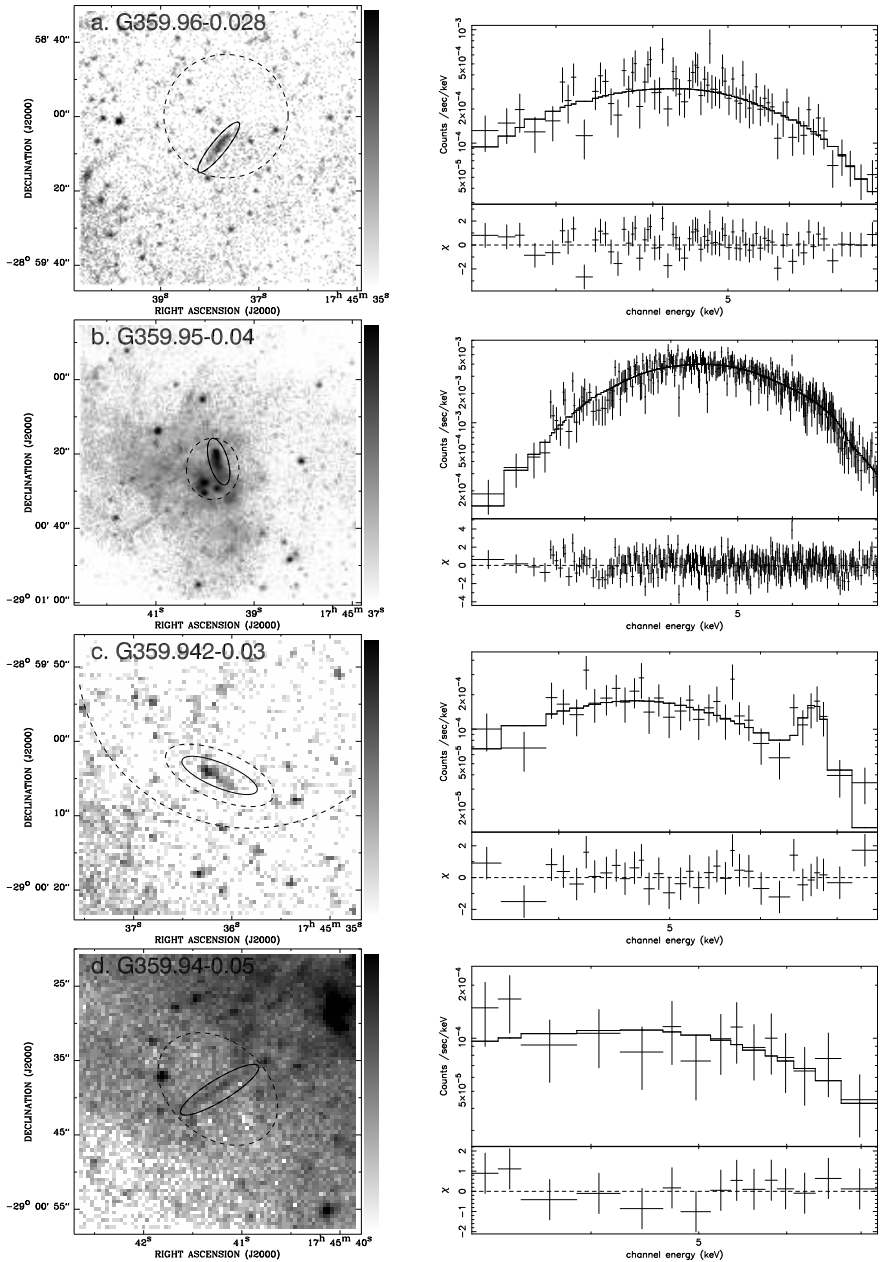

\include{fig5}
\caption{Continued X-ray count maps and spectra.  Upper and lower
  limits for each count map are 207 and 8.3
  counts/arcsec$^2$ for G359.96-0.028 (a), 1200 and 21
  counts/arcsec$^2$ for G359.95-0.04 (b), 207 and 12
  counts/arcsec$^2$ for G359.942-0.03 (c), and 410 and 4.1
  count/arcsec$^2$ for G359.94-0.05 (d), respectively. G359.942-0.03
  is fitted with a 6.7 keV Gaussian in addition to the \textsc{pha(po)} model.} 
\label{fig:fig5}
\end{figure*}

\begin{figure*}
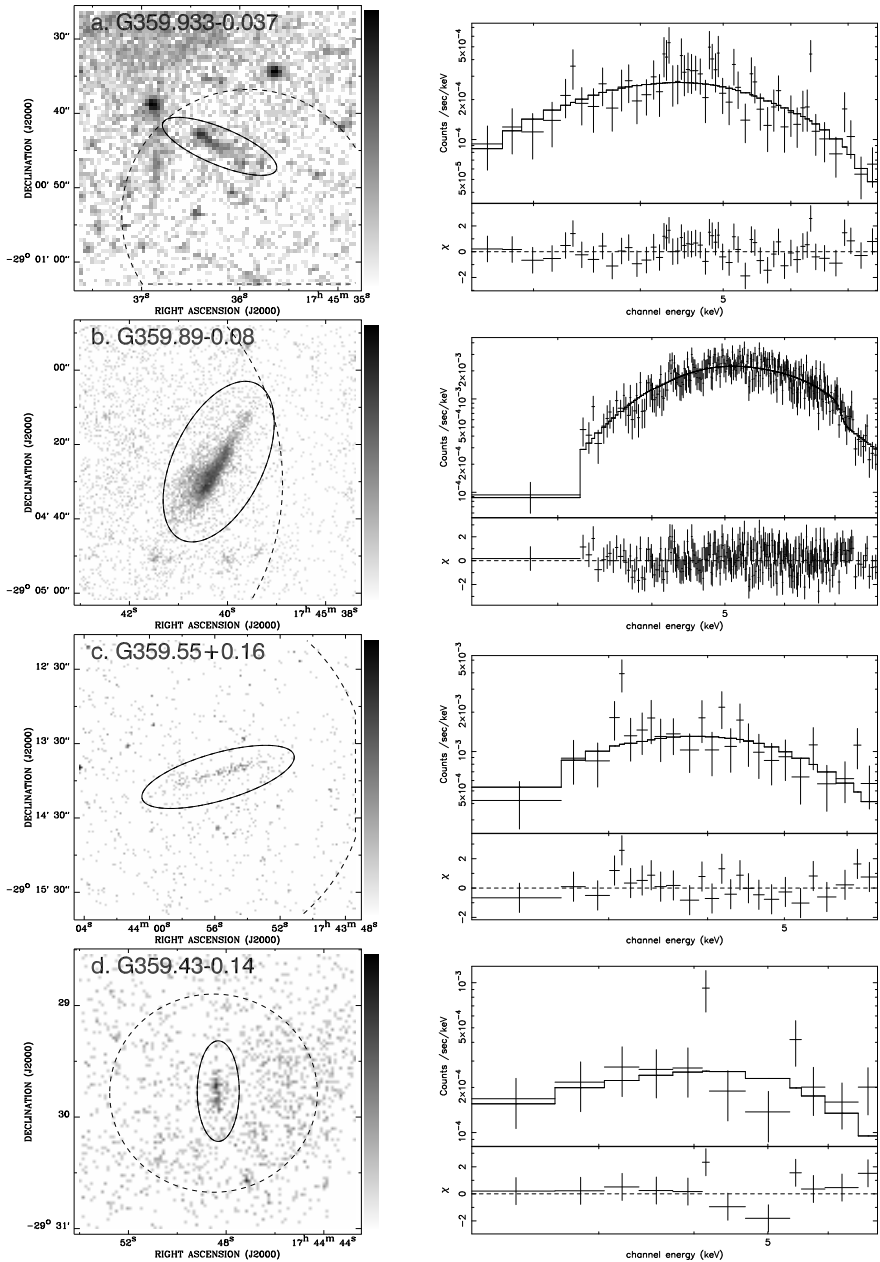

\include{fig6}
\caption{Continued X-ray count maps and spectra.  Upper and lower
  limits for each count map are 410 and 8.3
  counts/arcsec$^2$ for G359.933-0.037 (a), 410 and 4.1
  counts/arcsec$^2$ for G359.89-0.08 (b), and 12 and 0.5
  counts/arcsec$^2$ for G359.55+0.16 (c) and G359.43-0.14 (d), respectively.}
\label{fig:fig6}
\end{figure*}

\begin{figure*}
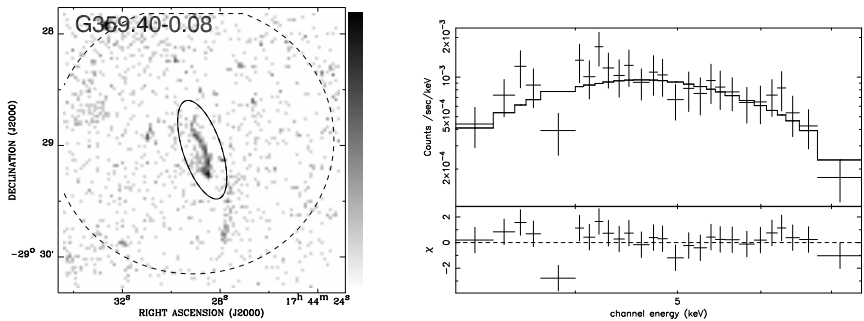

\include{fig7}
\caption{Final X-ray count map and spectra.  The count map
  has upper and lower limits of 12 and 0.5 counts/arcsec$^2$.}
\label{fig:fig7}
\end{figure*}

\begin{figure*}
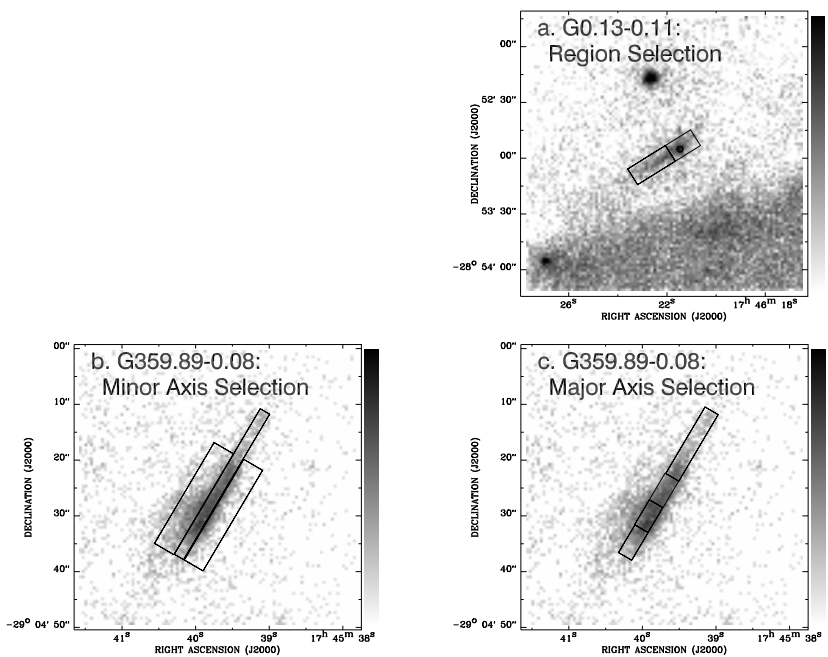

\include{fig8}
\caption{Region selection for piecewise analyses.  The count
  maps for G0.13-0.11 and G359.89-0.08 are overlaid with regions
  identifying the segments to be used in their respective piecewise
  analyses.  The region selection for G0.13-0.11 is shown in 8a.  The
  image 8b gives the minor axis selection for G359.89-0.08 while 8c
  gives the major axis selection. } 
\label{fig:fig8}
\end{figure*}

\subsection{G0.223-0.012}

G0.223-0.012 (Fig. 3a) is located near the bright X-ray point
source CXOGCS J174621.05-284343.2, removed from the image in Fig. 3a
with its location marked, to the northeast of the feature.  The
feature's position as well as orientation rejects the possibility that
G0.223-0.012 could be the result of a CCD readout streak from CXOGCS
J174621.05-284343.2. 

\subsection{G0.13-0.11}

G0.13-0.11 (Fig. 3b) was previously studied by \cite{wang02} as a possible
PWN.  Unlike other PWN candidates in the GC region, G0.13-0.11's
morphological proprieties are fairly unique in that it is curved
analogous to that of a co-existing non-thermal radio
filament in the Radio Arc region \citep[see][fig. 1]{wang02}.  The
morphology of G0.13-0.11 appears to include two tail-like segments
stemming from the apparent point source giving the feature a wing-like
appearance.  We identified the feature as cometary due to the more
prominent eastern segment.  \cite{wang02} proposed that the unique
morphology of this feature is likely due to interactions of
high-energy particles from the pulsar in a strong magnetic field,
traced by the radio polarization of the Radio Arc region.  To
determine if there is some spectral evolution across the feature, as
expected for PWNe, we perform a piecewise analysis of the apparent
point source and the 'wings' of G0.13-0.11 (see Fig. 8a).  In
separating the point source from the wing, we adopt the 1.5"
extraction radius used in \cite{wang02}.  A joint fit of the point
source and total wing spectra, keeping $N_{\rm{H}}$ a common parameter
between both features, produces a good fit ($\chi^2$/d.o.f=79.7/86) with
$N_{\rm{H}}$=6.0$^{+2.3}_{-3.6} \times 10^{22}$ cm$^{-2}$ and $\Gamma$ of
1.0$^{+.7}_{-.9}$ for the point source and 1.1$^{+0.3}_{-0.8}$ for
the wing; these values roughly agree with the analysis of
\cite{wang02}.  Segmenting the wing and performing the joint fit again
provides a joint fit column density of $N_{\rm{H}}$=5.2$^{+3.0}_{-2.6}
\times 10^{22}$ cm$^{-2}$ and $\Gamma$ of 0.9$^{+0.8}_{-0.7}$,
0.5$^{+0.6}_{-0.6}$, and 1.2$^{+0.7}_{-0.6}$ for the point source,
wing segment containing the point source and wing segment away from
the point source, respectively, with $\chi^2$/d.o.f=90.7/89.
These best fit spectral parameters determined from the joint fits are
in agreement with those listed in Table 2 which were determined for
the feature as a whole.  While the photon index appears to have
steepened away from the point source, the error bars for the photon
indexes of the two wing segments have considerable overlap.  As such,
we can not fully constrain the spectral evolution with the available
data.  

\subsection{G0.017-0.044}

Spectral analysis of G0.017-0.044 (Fig. 3d) shows a strong
6.4 keV emission line of neutral Fe K$\alpha$ transition.  Given its projected
separation from Sgr A*, approximately 11 pc, and this strong neutral
Fe emission, possible explanations for this feature include the
reflection of radiative illumination from Sgr A* past activity as
proposed by \cite{koyama89} and collisions in molecular clouds between
low energy cosmic-ray electrons and ions \citep[e.g.][]{valinia00},
though it is difficult to say with certainty due to the low photon
counts.  The relative flatness of the spectrum indicates that the
feature may result from a high temperature plasma.  This, however, is
inconsistent with the weakly ionized Fe responsible for the 6.4 keV
line unless the plasma is in a highly non-collisional ionization
equilibrium state.  \cite{lu08} studied a clump of diffuse X-ray 
emission located to the west of G0.03-0.06 (Fig. 3c) which
has similar line emission and equivalent width as G0.017-0.044.
This 'West Clump' \citep[as referred to by][]{lu08} is seen in the
images for G0.03-0.06 and G0.017-0.044 (Fig. 3c and 3d,
respectively).  In addition to the 6.4 keV line similarities, the
photon index of G0.017-0.044 is also comparable to that of the 'West
Clump' albeit with fewer photon counts and thus larger 90\% confidence
level error bars.  The similarities between the spectra of this West
Clump and G0.017-0.044 indicate that they may be produced through the
same mechanisms. 

\subsection{G359.942-0.03}

G359.942-0.03 (Fig. 5c) displays morphological properties similar to
a ram-pressure confined PWN but with a noticeable 6.7 keV emission
line from He-like Fe. The presence of this line and the steepness of
the spectrum indicates that the emission is likely thermal in nature,
contrary to the typical non-thermal PWN spectrum.  We thus apply the
\textsc{xspec} model \textsc{mekal} with absorption
(\textit{pha(mekal)}) for an optically-thin thermal plasma with
emission lines.  Using the default abundances (\textsc{xspec}
\textsc{angr}), the model provides a good fit ($\chi^2$/d.o.f=29.4/30)
with a temperature of 7.3$_{-2.3}^{+3.4}$ keV
($\sim$8.5$^{+3.9}_{-2.6}\times 10^7$ K) and
$N_{\rm{H}}$=18.6$^{+4.2}_{-3.8}\times 10^{22}$ cm$^{-2}$.  The best fit
normalization of 4.3$^{+1.2}_{-0.8}\times 10^{-5}$ gives a volume
emission measure of $\sim$1.1$^{+0.3}_{-0.2}$ cm$^{-6}$ pc$^3$. In
\S~4.2, we examine G359.942-0.03 as a ram-pressure confined stellar
wind bubble generated by a massive star.

\subsection{G359.89-0.08}

\cite{lu03} have previously studied G359.89-0.08 (Fig. 6b) in detail
from archived \textit{Chandra} data under the SNR and PWN scenarios.
The SNR case was determined unlikely in part due to a 10"-30" offset
between the non-thermal radio emission from the proposed SNR,
G359.92-0.09, and the X-ray emission from G359.89-0.08.  The high
X-ray absorption column density also questions the SNR scenario as it
suggests that G359.89-0.08 lies behind an enhancement of molecular
material, contrary to previous work on G359.92-0.09 which indicates
that it must lie, at least partially, in front of the molecular gas
\citep[see also][]{lu03,coil00}. In the previous study, they used two
observations pointed at Sgr A* (ID 242 and 1561) with an integrated
exposure time of 101 ks.  By performing a piecewise analysis of
G359.89-0.08, using the greater exposure time and counting statistics
of this study, we may be able to further constrain the physical model
of the X-ray emission.  We extract spectra from regions along the
major and minor axes (see Fig. 8b \& 8c) and perform power law
joint fits in which $N_{\rm{H}}$ is set as a common parameter for the
segments.  Examining the two fits, we see no noticeable trend across
the minor axis of the feature.  The outer regions have spectral
indexes of 1.3$^{+0.3}_{-0.4}$ and 1.5$^{+1.1}_{-1.1}$, for the SW and
NE segments, respectively, while the center segment has a spectral
index of 1.2$^{+0.3}_{-0.3}$, marginally harder then either of its
neighbors.  This is consistent with the PWN scenario for synchrotron
cooling of the pulsar wind materials; the outer regions probably
represent aging particles in a cocoon around a fresh stream from the
pulsar.  Across the major axis, we see evidence for possible spectral
hardening, or flattening, in G359.89-0.08 as shown in Fig. 9 and in
a contour plot of the feature (Fig. 10), though the overlap of the
90\% confidence level error bars prevents a definitive conclusion.  These
spectral trends do not provide supporting evidence for the SNR
scenario where one would expect any detectable systematic trend in the
spectral index to be along the minor axis while fairly uniform
along the major axis. Conversely, the apparent observed hardening
across the major axis is inconsistent with the softening expected for
the PWN scenario.  However, due to the relatively large uncertainty
present in the joint fits and the lack of a confirmed pulsar signal,
we are unable to judicially rule out the SNR case and constrain such
spectral evolution. 

\begin{figure}
\centerline{\includegraphics[height=.5\textwidth, angle=-90]{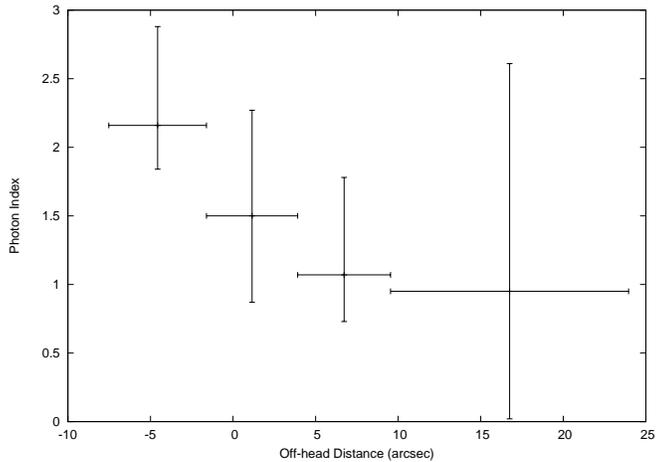}}
\caption[]{Plot of the photon index $\Gamma$ across the major axis of
  G359.89-0.08.  Positive distance is measured as going northwest away
  from the "head".  Error bars are given at 90\% confidence.}
\label{fig:fig9}
\end{figure}

\begin{figure}
\centerline{\includegraphics[width=.5\textwidth]{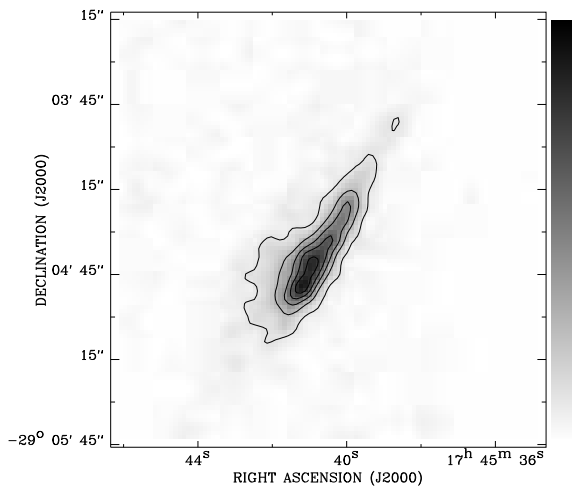}}
\caption{Contour plot of G359.89-0.08.  The image, bin size of 1
  pixel ($\sim$0.5"), was smoothed using a Gaussian with a FWHM of 3
  pixels ($\sim$1.5").  The upper and lower limits were placed at 124
  and 4.1 counts/arcsec$^2$, respectively.  The contour levels were
  placed with a step size of 21 counts/arcsec$^2$ and a starting value
  of 17 counts/arcsec$^2$. }
\label{fig:fig10}
\end{figure}

\subsection{Joint Analysis of Faint Features}

The features G0.223-0.012, G359.55+0.16, G359.43-0.14 and G359.40-0.08
(Fig. 3a, 6c, 6d and 7, respectively) were observed over much lower
effective exposure times (in the range
of $\approx$100 to 150 ks) than those features near the
central region ($\approx$700 ks to 1 Ms).  Due to the lower
signal-to-noise of these features, the inferred model parameters tend to have
rather large 90\% confidence level error bars (e.g. $N_{\rm{H}}$ upper-lower
bound difference of $>10^{23}$ and $\Gamma$ upper-lower bound
difference of $>2$). To better constrain the parameter bounds, we
perform joint fits of these 4 'faint' features (faint in terms of low
signal-to-noise ratios as their fluxes are still fairly high to be
seen in the low exposure time).  In performing the joint fits,
we fixed $\Gamma$ or $N_{\rm{H}}$ to be the same for all features while
allowing it and the remaining parameters to vary for the fit;
e.g. fixing $N_{\rm{H}}$ to be a common parameter between all features and
running the fit over the common $N_{\rm{H}}$ and each feature's $\Gamma$.
From these joint fits, we find that the spectral parameters of the
'faint' features are consistent over both fits.  That is, the $N_{\rm{H}}$
of the individual features in the $\Gamma$ joint fit is consistent
with the average $N_{\rm{H}}$ from the $N_{\rm{H}}$ joint fit and similarly for
$\Gamma$.  The spectral parameters for the $N_{\rm{H}}$ joint fit are given
at the bottom of Table 2.  Due to the low signal-to-noise ratios of
the individual features, constraining their spectra and morphology is
difficult, even with the joint fits.  Nevertheless, these fits show
the relative flatness present between the features and the overall
photon index consistent with a primarily non-thermal origin.

\section{Nature of the Filamentary X-ray Features}

\subsection{Comparison of GC Linear X-ray Threads and Known Pulsars}

Outside of the GC, similar linear X-ray features are found directly
associated with known pulsars, providing the basis for our PWN
interpretation.  \cite{karga08} have collected a number of PWNe detected
with \textit{Chandra}.  Many of the PWNe listed are affected by their relative
motion with the ambient medium, e.g. ram-pressure confinement.  These
features closely resemble those in the GC in their morphological
appearance, most notably the Mouse PWN \citep[ID 22 in][]{karga08}.
Indeed, a number of the features presented in fig. 3 of
\cite{karga08} would appear similar to those found in the GC including
pulsars J1809-1917, J1509-5850, B1853+01, G189.23+2.90, G326.12-1.81,
and G327.15-1.04 \citep[ID 26, 29, 30, 47, 51, 52;][]{karga08}. It
should be noted that not all X-ray features associated with pulsars
can be easily classified using a simple model, such as the peculiar
case of the pulsar B2224+65, also responsible for the Guitar Nebula,
and its linear X-ray feature which is $\sim$118$^\circ$ offset from
the pulsar's proper motion \citep{hui07,johnson09}.  

Based on known pulsar/X-ray feature associations, we are able to
derive simple models for many of the GC X-ray filaments.  While the
X-ray counting statistics for many of the features studied are indeed low, the
power law model provides a reasonable fit to the spectra.  Out of the
17 features presented, we identify 7 as PWNe based on their elongated
morphology and spectral parameters.  The features G0.13-0.11,
G0.03-0.06, G359.97-0.038, G359.97-0.009, G359.964-0.052,
G359.95-0.04, and G359.89-0.08 have spectral indexes in the range of
1.1 to 1.9 with 2-10 keV luminosities of 5$\times$10$^{32}$ to
$\sim$10$^{34}$ ergs s$^{-1}$, consistent with typical PWNe
\citep[e.g.][]{gotthelf02,gaensler06,kaspi06}.  The spectral
parameters of these features support the PWN scenario, though the SNR
case can not be ruled out, while the low signal-to-noise ratios of
many of the other features do not give enough information to
accurately constrain their physical models.  Moreover, scattering
towards the GC makes the detection of the radio pulsar signal
difficult, preventing a definite confirmation of the PWN scenario at
present.  Detection of pulsars in the X-ray band depends on the
pulsar's orientation and mechanism for producing X-ray
photons. Additionally, we currently lack the necessary instrumentation
with enough sensitivity to perform X-ray timing analysis of the
putative pulsars with the observed faint X-ray fluxes. 

For those features that show emission lines (namely G0.017-0.044 
and G359.942-0.03), the inclusion of a Gaussian profile with the power-law
model reduces the $\chi^2$ compared to the power-law only fit with F-test
significance $>$98\%.  It should be noted that, while common practice,
using the F-test to test for the presence of line emission has a
number of short comings \citep[see][]{protassov02}.  However, we are
unaware of a simple, yet statistically rigorous, approach that would
better determine the significance of any lines present.
Statistically, thermal models, such as Bremsstrahlung, produce
satisfactory fits but would have unreasonably high temperatures,
i.e. $>$10 keV \citep[see also][]{lu08}.  The presence of lines at
either 6.4 keV or 6.7 keV are obvious indicators that the features are
not PWNe.  The 6.4 keV line, in particular, is likely due to
fluorescent emission from either reflection of transient bright X-ray
sources (e.g. Sgr A*; \citet{koyama89}) or local enhanced low energy
cosmic ray electrons \citep{yusef02}.  Similar line emissions have
been seen in the Galactic Ridge; however, the 6.4 keV line emission in
the ridge is relatively weak, compared to other lines such as the 6.7
keV transition, and is consistent with an origin in numerous
point-like sources \citep[most likely cataclysmic
  variables,][]{ebisawa08,revnivtsev06}.

\subsection{G359.942-0.03 as a Ram-Pressure Confined Stellar Wind Bubble}

We find a point-like near-IR source that may be associated with
G359.942-0.03 (Fig. 5c) in the 2MASS catalog \citep[ID
  17453582-2900050,][]{skrutskie06} and in a recent HST/NICMOS survey
of the GC at 1.90 $\mu$m \citep{wang09}. The 2MASS counterpart has an 
apparent H-band magnitude of 10.85$\pm0.06$ and an H-K color of 1.67 mag, 
in agreement with those observed for massive stars in the GC
region.  The 2MASS source has an angular separation of .71" with
respect to the approximate center of the "head" of G359.942-0.03.
Within 30" of G359.942-0.03, there are a total of 48 2MASS sources
relating to a P-statistic (chance of random association, calculated
using P=1-e$^{-\pi n\theta^2}$ where n is the 2MASS source density and
$\theta$ is the distance between potential counterparts, see also
\cite{borys04,pope06}) of $\sim$3\% for the potential 2MASS
counterpart.  If one considers the size of the extended X-ray
emission, then this P-statistic increases to $\sim$50\%; however, if
this model is correct, we should only consider the region where the
massive star would be located; e.g. in the "head", leading to a robust
identification for the 2MASS counterpart.  Interestingly,
the color excess E(H-K)=1.72 (assuming (H-K)$_{intrinsic}\approx$-0.05
for early-type stars \citep[see also][]{figer99a,panagia73}) is
significantly lower than that inferred from the X-ray absorption. This
discrepancy is expected if the star has a strong wind, such as a
Wolf-Rayet star, which can produce a free-free emission enhancement
along with H/He lines in the H and K bands \citep[see
  also][]{figer97,figer99a,figer02}.  As an example, the brightest
massive star identified in the Arches Cluster \citep{figer02} (2MASS
ID 17455043-2849215) has apparent H and K band magnitudes of 9.5 and
8.6, respectively, giving E(H-K)=0.95.  Additional redding may come
from the large variations in extinction around the GC over large
distances.  If we can then assume that this source is a massive star
with strong winds, one may expect to see an extended emission
region in the Pa$\alpha$ map of \cite{wang09}; however, the expected
extinction based on the X-ray column density would place the
Pa$\alpha$ flux well below the detection limit of the survey.  This
indicates that while the presence of strong winds associated with a
massive star can not be confirmed, this scenario remains plausible for
the 2MASS counterpart of G359.942-0.03.

This ram-pressure confined wind bubble model is driven by the presence
of the massive star as well as the cometary morphology and the
apparent thermal spectrum of the X-ray emission (\S~3.4).  In Fig.
5c, we can see fairly well constrained extended X-ray emission, with
apparent dimensions of $\sim$2"$\times$6" or .08$\times$.23 pc, with a
bright "head" region, consistent with our definition of a cometary
feature.  In this model, the motion of the star creates a bow shock
ahead of it that sweeps up material from the ISM, similar to the model
presented by \cite{vanburen88}.  We may infer a number of model
parameters for this scenario based on the properties of the X-ray-emitting gas
(\S~3.4). First, the strong stellar wind from the massive star is
expected to be heated in a reverse shock to a temperature of $T \sim 6
\times 10^7 {\rm~K}  (v_{\rm{w}}/2)^2$, where $v_{\rm{w}}$ is the stellar wind
velocity (in units of $10^3$ km s$^{-1}$).  The possible range of
stellar wind velocities under the massive star scenario, typically of
1-2$\times 10^3$ km s$^{-1}$, can then produce temperatures high
enough to match the value ($\sim$8.5$^{+3.9}_{-2.6}\times 10^7$ K) inferred
from the spectral analysis.  Second, the measured volume emission 
measure ($\sim$1.1$^{+.3}_{-.2}$ cm$^{-6}$ pc$^3$) provides an
estimate on the density of the shocked wind material of 
$n_e \sim$ 30 cm$^{-3}$, assuming a cylindrical volume of the observed
width and length (Table 1). Third, constraints may be placed on the
stellar velocity from balancing the ram and shocked wind pressures,
$2 n_e k T \sim 10^{10}\mu_{\rm{H}} m_{\rm{H}} n_0v_*^2$, where $v_*$ is the 
stellar velocity (km s$^{-1}$), $\mu_{\rm{H}}$ is the average molecular
weight per hydrogen atom ($\mu_{\rm{H}}$=0.6 for approximately 
solar metallicity) and $n_0$ is the density of the ambient 
ISM (cm$^{-3}$).  This balance of the ram-pressure also serves to
determine the size of the bow shock region.  Van Buren \& McCray
(1988) showed that the distance to the contact discontinuity of the
bow shock is $\sim 1.5l_1$ where $l_1$ is the length scale
characterizing the point where the ISM ram pressure equals that of the
stellar wind.  The length scale $l_1$ is then given by 
\begin{equation}
l_1=(56 {\rm~pc}) \dot{M}^{1/2}v_{\rm{w}}^{1/2}v_*^{-1}\mu_{\rm{H}}^{-1/2}n_0^{-1/2}
\end{equation}
where $\dot{M}$ is the mass loss rate in units of $10^{-6}$ M$_{\sun}$
yr$^{-1}$ \citep[see also][eq. 1]{vanburen88}. Since the measured
width of G359.942-0.03, $\sim$0.08 pc taken across the "head" region,
should be $\sim 3l_1$, as the shocked wind gas is primarily
responsible for the X-ray emission, we have a stellar velocity
estimate of
\begin{equation}
v_* \sim (27 {\rm~km~s^{-1}}) \dot{M}^{1/2}v_{\rm{w}}^{1/2} (n_0/10^4)^{-1/2}. 
\end{equation}

From the above equations, we then infer $v_*^2 n_0 \sim 1.5-7\times
10^7$ assuming $\dot{M}=1$ and $v_{\rm{w}}=2$ or $v_*\sim 39-84$ km
s$^{-1}$ for $n_0\approx 10^4$ cm$^{-3}$, consistent with what may be expected
from a run-away massive star into a dense cloud.  Based on typical
parameters for massive stars, this ram-pressure confined wind bubble
model is capable of explaining the thermal spectrum, luminosity and
well collimated, comet-like morphology of the X-ray feature
G359.942-0.03 along with its apparent association with a massive star.
We should caution readers that this conclusion, while reasonable, is
not definite due to the number of assumed parameters.  Further
analysis, including deeper infrared imaging and spectroscopic
observations, can aid in determining the assumed values and provide
confirmation (or contradiction) to this ram-pressure confined stellar
wind bubble model. 

\section{Summary And Final Remarks}

To conclude, X-ray features in the GC likely have a heterogeneous origin.
Non-thermal features are typically produced through synchrotron emission
of PWNe or magnetohydrodynamical shock fronts from SNRs.  For the PWN
model, one would expect elongation due to either the pulsar movement
in the ISM or magnetic field confined flows of the pulsar wind
materials.  In such a case, one would expect to see spectral softening
across the direction of elongation along with typical photon indexes of
$\Gamma$=1.1-2.4.  For the SNR case, the direction of elongation would be
caused by shock fronts and any measurable spectral change would be
expected across the width of the feature.  Of the features presented,
7 exhibit characteristics of PWN, although the SNR case has not been
completely ruled out.  Their non-thermal spectra as well as typical
comet-like morphology are consistent with the ram-pressure confined
PWN model.  The lack of any detected pulsar signal from these PWNe
prevents a definitive conclusion on their origin.  Some of these
pulsars may be detected in future pulsar searches using high frequency
and high resolution radio observations.  In the case of G359.942-0.03,
the model of a ram-pressure confined stellar wind bubble best explains
the detection of the 6.7 keV He-like Fe emission and steep spectral shape.
The near-IR images show an apparent counterpart of G359.942-0.03
consistent with a massive star where the lack of extended near-IR emission
could be due to a combination of extinction and inefficiency in
ionizing photon conversion.

In addition to their physical models, the X-ray features provide
information into the underlying structure of the GC.  Features such as
G0.017-0.044 and the "clumps" as studied by \cite{lu08} map out the
history of the GC (e.g. the possible radiative illumination of
G0.017-0.044 from Sgr A*).  The orientations of PWN-like
features help to map the magnetic, gas and stellar dynamics of the GC.
\cite{lu08} predicted that the orientation of comet like features is
due to a strong galactic nuclear wind comparable to typical pulsar
velocities oriented away from Sgr A*.  Of the features present within
a 20'$\times$20' image centered on Sgr A* (see Fig. 2), this effect
can be seen in 5 out of the 7 PWN candidates (excluding G359.95-0.04
and G359.89-0.08) and G359.933-0.037.  Those features outside the
20'$\times$20' region along with the features G359.964-0.03,
G359.95-0.04 and G359.89-0.08 do not necessarily follow this claim;
though this could be explained by variations in the vector field of
the galactic wind as a result of regions of varying density.
Additionally, the filament-like features could be regulated by local 
ordered magnetic field lines as with G0.13-0.11's morphology.  These
features can thus be used in more in-depth studies of GC dynamics,
beyond the quick analysis just presented, in combination with deep
observations at other wavelengths; e.g. radio polarization and
magnetic field tracers.

For the GC observations, few X-ray filaments were identified in
regions outside the central 20'$\times$20' region.  These regions have
integrated exposure times significantly less than the central region.
In addition, few of the identified features within the central
20'$\times$20' region were found in initial observations with exposure
times $<$200 ks.  Thus, those presented in this study may represent only a
tip of the "iceberg" of X-ray features.  Indeed, some may be
aligned such that they appear like point sources with their direction
of elongation along the line of sight.  Based on the average star
formation rate of the GC \citep[$\sim 10^{-7}$ M$_{\sun}$ yr$^{-1}$
  pc$^{-3}$,][]{figer04}, \cite{lu08} predicted $\sim$40 pulsars
within 7 pc of the GC with ages $\lesssim 3\times 10^5$ yr assuming an
average stellar mass of 10$M_{\sun}$.  This is roughly consistent with
our 13 PWN candidates if we consider that projection effects will
limit us to $\sim$71\% of PWN, assuming the extended X-ray emission of
a PWN is identified if the angle between the extended emission and the
line of sight lies between $\pi$/4 and 3$\pi$/4, along with factors
including sensitivity limits and pulsar population statistics.  With
higher sensitivity and spatial resolution in multiple wavelength bands,
including X-ray, radio and near-IR, the identities of these X-ray
features can be more accurately determined.

\section*{Acknowledgments}

The project was funded in part by the North East Alliance and
the Nation Science Foundation under grant number NSF HRD 0450339 and
by NASA/SAO through grant GO7-8091B.  This publication makes use of
data products from the Two Micron All Sky Survey, which is a joint
project of the University of Massachusetts and the Infrared Processing
and Analysis Center/California Institute of Technology, funded by the
National Aeronautics and Space Administration and the National Science
Foundation.  We would also like to thank referee Pat Slane for his comments
on improving the manuscript.


\clearpage

\begin{deluxetable}{llccc}
\tabletypesize{\scriptsize}
\tablewidth{0pt}
\tablecaption{Morphological Properties of Identified X-ray Features}
\tablehead{\colhead{ID}& \colhead{Shape}& \colhead{Size}& \colhead{Orientation}& \colhead{Notes}\\
& & pc & &}

\startdata
G0.223-0.012&	filamentary&	0.23$\times$3.12&	ESE-WNW& Faint\\
G0.13-0.11&	cometary&	0.28$\times$1.29&      ESE-WNW& Curved
feature\\
G0.03-0.06&	cometary&	0.27$\times$1.16&	SE-NW& Curved 'tail'\\
G0.017-0.044&	filamentary&	0.08$\times$0.50&	ESE-WNW& Faint\\
G0.007-0.014&	filamentary&	0.12$\times$0.39&	ENE-WSW& Faint\\
G359.97-0.038&	cometary&	0.23$\times$0.54&	SW-NE&\\
G359.97-0.009&	cometary&	0.08$\times$0.39&	SSE-NNW&\\
G359.964-0.052&	filamentary&	0.08$\times$0.50&	NNE-SSW&\\
G359.96-0.028&	filamentary&	0.12$\times$0.39&	SE-NW&\\
G359.95-0.04&	cometary&	0.08$\times$0.31&	NNE-SSW&\\
G359.942-0.03&	cometary&	0.08$\times$0.23&	ENE-WSW& Faint\\
G359.94-0.05&	filamentary&	0.08$\times$0.39&	ESE-NWN&\\
G359.933-0.037&	cometary&	0.08$\times$0.27&	ENE-NWN& Faint\\
G359.89-0.08&	cometary&	0.31$\times$0.93&	SE-NW&\\
G359.55+0.16&	filamentary&	0.31$\times$2.18&	E-W& Faint\\
G359.43-0.14&	cometary&	0.15$\times$0.83&	N-S& Faint\\
G359.40-0.08&	cometary&	0.20$\times$1.07&	S-N& Faint, Curved feature\\

\enddata
\tablecomments{\footnotesize{Sizes are taken from apparent surface
    brightness as compared to the local background.  Orientation is based
    on North and East defined as up and left in \textit{Chandra} images,
    respectively.  Notation for orientation then follows that of the
    cardinal directions.  "Faint" refers to features that have low
    signal-to-noise or low surface brightness when compared to the
    respective local background.}}
\end{deluxetable}

\begin{deluxetable}{lllccccr}
\tabletypesize{\scriptsize}
\tablewidth{0pt}
\tablecaption{X-ray Spectral Properties of Identified Features}
\tablehead{\colhead{ID}& \colhead{$N_{\rm{H}}$} & \colhead{$\Gamma$} & \colhead{$F_{\rm{x}}$} & \colhead{$L_{\rm{x}}$}  & \colhead{EW} & \colhead{Line Energy} & \colhead{$\chi^2/\nu$}\\
& (10$^{22}$ cm$^{-2}$) & & (10$^{-14}$ ergs/s/cm$^2$)&  (10$^{32}$ ergs/s)& (keV) & (keV)}
\startdata
G0.223-0.012&	5.1$_{-5.1}^{+79.8}$&	0.6$_{-1.6}^{+9.4}$&	15&	15&	\nodata&	\nodata&	6.7/8\\
G0.13-0.11&	7.0$_{-1.9}^{+2.2}$& 1.52$_{-0.2}^{+0.5}$&	34&	40&	\nodata& \nodata& 96.0/100\\
G0.03-0.06&	6.3$_{-2.8}^{+1.9}$&	1.1$_{-0.4}^{+0.4}$&	12&	13&	\nodata&	\nodata&	118.9/119\\
G0.017-0.044&	0.0$_{-0.0}^{+22}$&	        -0.7$_{-0.6}^{+3.0}$&	4&	3&	0.62$_{-0.34}^{+0.58}$&	6.4$_{-0.1}^{+0.1}$&	15.4/16\\
G0.007-0.014&	5.7$_{-5.7}^{+16.8}$&	1.0$_{-1.7}^{+3.4}$&	2&	2&	\nodata&	\nodata&	24.3/20\\
G359.97-0.038&	11.7$_{-1.9}^{+5.1}$&	1.4$_{-0.3}^{+0.8}$&	11&	15&	\nodata&	\nodata&	109.7/137\\
G359.97-0.009&	9.6$_{-5.4}^{+8.8}$&	1.2$_{-0.6}^{+0.7}$&	4&	5&	\nodata&	\nodata&	52.5/59\\
G359.964-0.052&	11.1$_{-1.5}^{+2.4}$&	1.9$_{-0.3}^{+0.5}$&	23&	37&	\nodata&	\nodata&	148.4/177\\
G359.96-0.028&	7.2$_{-3.2}^{+4.5}$&	0.9$_{-0.4}^{+0.6}$&	5&	5&	\nodata&	\nodata&	66.3/68\\
G359.95-0.04&	6.0$_{-1.0}^{+2.0}$&	1.8$_{-0.2}^{+0.3}$&	52&	93&	\nodata&	\nodata&	300.7/304\\
G359.942-0.03&	58.8$_{-32.8}^{+32.8}$&	4.1$_{-2.4}^{+3.0}$&	2&	20&	0.82$_{-0.22}^{+0.46}$&	6.7$_{-0.1}^{+0.1}$&	22.3/28\\
G359.94-0.05&	-0.0$_{-0.0}^{+9.3}$&	-0.4$_{-1.0}^{+1.4}$&	3&	2&	\nodata&	&	5/12\\
G359.933-0.037&	8.3$_{-4.2}^{+5.8}$&	0.7$_{-0.5}^{+0.4}$&	5&	5&	\nodata&	\nodata&	45.3/53\\
G359.89-0.08&	31.4$_{-2.0}^{+5.4}$&	1.3$_{-0.2}^{+0.6}$&	47&	104&	\nodata&	\nodata&	246.9/240\\
G359.55+0.16&	5.3$_{-3.5}^{+7.1}$&	1.1$_{-0.6}^{+1.9}$&	19&	20&	\nodata&	\nodata&	20.7/23\\
G359.43-0.14&	0.0$_{-0.0}^{+4.5}$&	-0.4$_{-0.7}^{+1.5}$&	6&	5&	\nodata&	\nodata&	15.2/9\\
G359.40-0.08&	16.2$_{-10.4}^{+12.6}$&	1.3$_{-1.4}^{+1.7}$&	15&	24&	\nodata&	\nodata&	22.1/24\\
"Faint" Features& 5.5$_{-3.6}^{+3.9}$&\multicolumn{5}{c}{\nodata}& 70.3/67\\
G0.223-0.012	&\nodata       &0.7$_{-1.4}^{+1.3}$&            15&   15\\
G359.55+0.16	&\nodata       &1.2$_{-1.1}^{+1.3}$&	        18&	20\\
G359.43-0.14	&\nodata       &1.1$_{-1.3}^{+1.0}$&            4&      4\\
G359.40-0.08	&\nodata       &-0.1$_{-0.6}^{+0.7}$&           20&	18
\enddata
\tablecomments{\footnotesize{Spectral parameters obtained through
    \textsc{xspec} model fitting (\textsc{pha(po)} or \textsc{pha(po+ga)}) to
    each of the identified X-ray features. In order,
    the parameters are the hydrogen gas column density $N_{\rm{H}}$, X-ray
    photon index $\Gamma$, observed X-ray flux and unabsorbed X-ray
    luminosity in the 2.0-10.0 keV band $F_{\rm{X}}$ and $L_{\rm{X}}$,
    equivalent width and energy centroid of the emission line if present, and
    $\chi^2$ per degree of freedom.  Luminosities are based
    on the assumed distance of 8 kpc to the Galactic center. Spectral
    parameters for "faint" features are based on the joint fit model
    with $N_{\rm{H}}$ as a common parameter. Errors are given to the 90\%
    confidence interval.}}
\end{deluxetable}

\end{document}